\documentclass[12pt]{iopart}

\usepackage{bm}
\usepackage{graphicx}

\usepackage{latexsym}
\usepackage{amssymb}
\expandafter\let\csname equation*\endcsname\relax 
\expandafter\let\csname endequation*\endcsname\relax
\usepackage{amsmath}
\usepackage{xcolor}
\usepackage{cite}
\usepackage[normalem]{ulem}

\def\be{\begin{equation}}
\def\ee{\end{equation}}
\def\bea{\begin{eqnarray}}
\def\eea{\end{eqnarray}}
\def\ra{\rangle}
\def\la{\langle}
\def\bi{\begin{itemize}}
\def\ei{\end{itemize}}
\def\ben{\begin{enumerate}}
\def\een{\end{enumerate}}

\DeclareMathOperator*{\argmax}{arg\,max}

\usepackage{color}

\begin{document}

\title{Phase diagram and optimal control for $n$-tupling discrete time crystal}

\author{Arkadiusz Kuro\'s\textsuperscript{1}, Rick~Mukherjee\textsuperscript{2}, Weronika Golletz\textsuperscript{1}, Frederic Sauvage\textsuperscript{2}, Krzysztof Giergiel\textsuperscript{1}, Florian Mintert\textsuperscript{2} and  Krzysztof Sacha\textsuperscript{1} }

\address{\textsuperscript{1} Instytut Fizyki Teoretycznej, 
Uniwersytet Jagiello\'nski, ulica Profesora Stanis\l{}awa \L{}ojasiewicza 11, PL-30-348 Krak\'ow, Poland}
\address{\textsuperscript{2} Blackett Laboratory, Imperial College London, SW7 2AZ, UK}
\ead{arkadiusz.kuros@uj.edu.pl}
\vspace{10pt}
\begin{indented}
\item[]
\end{indented}

\begin{abstract}
A remarkable consequence of spontaneously breaking the time translational symmetry in a system, is the emergence of time crystals. In periodically driven systems, discrete time crystals (DTC) can be realized which have a periodicity that is $n$ times the driving period. However, all of the experimental observations have been performed for period-doubling and period-tripling discrete time crystals. Novel physics can arise by simulating many-body physics in the time domain, which would require a genuine realisation of the $n$-tupling DTC. A system of ultra-cold bosonic atoms bouncing resonantly on an oscillating mirror is one of the models that can realise large period DTC. The preparation of DTC demands control in creating the initial distribution of the ultra-cold bosonic atoms along with the mirror frequency. In this work, we demonstrate that such DTC is robust against perturbations to the initial distribution of atoms. We show how Bayesian methods can be used to enhance control in the preparation of the initial state as well as to efficiently calculate the phase diagram for such a model. Moreover, we examine the stability of DTCs by analyzing quantum many-body fluctuations and show that they do not reveal signatures of heating.
\end{abstract}

\section{Introduction} 
Ever since the original conception of a time crystal in quantum many-body systems \cite{wilczek2012}, there has been a growing interest to understand these objects theoretically \cite{Sacha2015,PhysRevLett.116.250401,PhysRevLett.117.090402,Yao2017,Russomanno2017,Gong2017,Huang2017,Iemini2017,Zeng2017,Sacha2018,Sacha_rew,Mizuta2018,Liao2018,Lesanovsky2019,khemani2019brief, Cosme2019} as well as to realise them experimentally \cite{Monroe2017,choi2017observation,Pal2018,Rovny2018,Rovny2018a,Smits2018}. It turns out that in the model proposed in \cite{wilczek2012} continuous time translation symmetry breaking cannot be observed if a system is initially in the ground state \cite{Bruno2013}, while excited-state realizations are allowed \cite{syrwid2017}. Furthermore, time crystals cannot be observed for systems in the presence of long-range power-law interaction in their equilibrium state \cite{Watanabe2015,Watanabe2019}. However, recent study shows~\cite{Kozin2019} that the time crystal behaviour can also be observed in the ground state of a system with long-range interactions in the form of spin strings which are hard to realize in real experiments \cite{Khemani2020comment}. Popular realisations of time crystals involving excited states are DTCs which arise in periodically driven many-body systems. In such systems, the discrete temporal symmetry can be broken as a result of the inter-particle interactions. The first proposal of DTC involved a bouncing gas of ultra-cold atoms \cite{Sacha2015} and latter proposals involved spin-1/2 systems \cite{PhysRevLett.116.250401, PhysRevLett.117.090402}. So far, the experimental realization of DTC has been performed in trapped ions \cite{Monroe2017} and nitrogen-vacancy centres in a diamond \cite{choi2017observation}, where period-doubling DTC and period-tripling DTC was observed respectively (see also \cite{Pal2018,Rovny2018,Rovny2018a,Smits2018}). Although there are few theoretical models to realize period $n$-tupling DTC \cite{Sacha2018,Surace2018,PhysRevA.99.033626,pizzi2019,Pizzi2019a}, where $n>3$, there has yet been no experimental observation of them.

In this work, we study a model of ultra-cold bosonic atoms bouncing resonantly on an oscillating mirror \cite{Sacha2015} which has the potential of realizing $n$-tupling DTC, where $n$ can be arbitrarily large. The motivation to physically realise $n$-tupling DTC lies in the fact that they provide a suitable platform to exhibit topological time crystals \cite{Lustig2018,Giergiel_2019}, temporal quasi-crystals \cite{Li2012,Huang2017a,PhysRevLett.120.140401,PhysRevB.99.220303,Pizzi2019a} as well as to demonstrate various nontrivial condensed-matter phenomena in the time domain \cite{sacha2015anderson,PhysRevA.94.023633,Sacha2018,PhysRevLett.119.230404,PhysRevB.96.140201,PhysRevA.97.053621,PhysRevA.98.023612}. However, the experimental realization of this model involves multiple challenges as detailed in \cite{Sacha2018,Giergiel2020}.

The formation of DTC in the system occurs due to sufficiently strong attractive interaction between atoms and it can be observed if an initial state is properly located with respect to the mirror position \cite{Sacha2018}. This requires a precise system control which cannot be guaranteed in each experimental realization. By evaluating the phase diagrams of this model, we examine the robustness of the DTC against perturbations of the initial atomic distribution. Moreover, we theoretically investigate the possibility of applying optimal control based on Bayesian optimization which could be performed on the experimental realisations of DTCs. We also address the question of whether the DTCs are stable against quantum many-body fluctuations which can result in heating of the system. Although the model we study allows arbitrarily large $n$, as a proof of principle, thorough analysis for period-doubling and period-quadrupling DTCs are addressed in this work. This will be hugely beneficial for real experiments which will most likely be carried out with a larger $n$ to reduce possible atom losses. 

The paper is organized as follows. In Section \ref{system} we present the system of ultra-cold bosonic atoms bouncing on an oscillating atom mirror. The results concerning the phase diagrams and optimal control are shown in Section~\ref{ph_diag} and Section~\ref{opt}, respectively, while quantum many-body fluctuations of DTCs is presented in Section~\ref{Bogoliubov}.

\section{Theory: Model for period $n$-tupling discrete time crystal} \label{system}
We provide here only a brief description of the model of the $n$-tupling DTC in ultra-cold atoms, the full description in all its details can be found in Refs.~\cite{Sacha2015,Sacha2018,Giergiel2020}.

\begin{figure}[ht!] 	    
\centering        
\includegraphics[width=0.7\columnwidth]{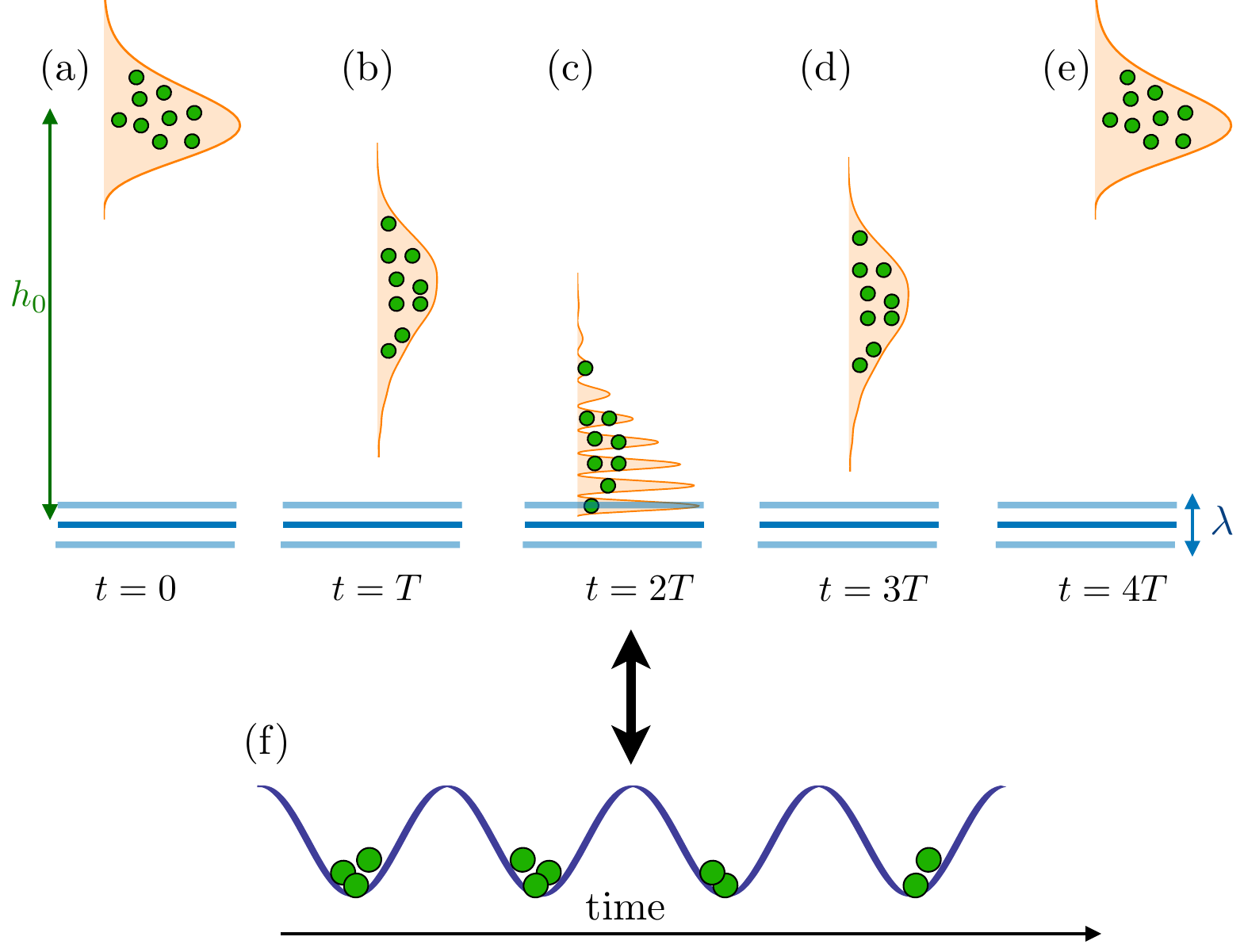}                
\caption{Setup figure of bouncing atoms on an oscillating mirror with period $T$ for an $n:1$ resonant driving where $n=4$. (a) The initially prepared atomic distribution at the classical turning point $h_0$ returns to the starting position at $t=4T$ (e). Figures (b)-(d) show the atomic distribution for time $t=T$, $t=2T$ and $t=3T$, respectively. (f) In the effective description, the system is described by Bose-Hubbard model, see~(\ref{ener1}).} 
\label{setup}   
\end{figure}

We consider $N$ ultra-cold bosonic atoms which are bouncing resonantly on an oscillating atom mirror. We assume that the system is strongly confined in the transverse directions and can be treated within the one-dimensional approximation. Atoms interact with each other via a contact potential whose strength $g_0$ is proportional to the s-wave scattering length $a_s$ and can be controlled by means of the Feshbach resonance mechanism \cite{Pethick2002}. We restrict the discussion to attractive interactions $g_0<0$. The system can be effectively described by a Bose-Hubbard model, Eq.~(\ref{ener1}), which is schematically presented in Fig. \ref{setup}f. It turns out that for sufficiently strong attractive interactions, the system spontaneously breaks discrete time translation symmetry of the Hamiltonian and starts evolving with a period $n$-times longer than the period of the Hamiltonian (see  Fig. \ref{setup}a-e). Let us stress that the attractive interactions between atoms do not cause a collapse of an atomic cloud in the one-dimensional space and atoms are able to form a Bose-Einstein condensate (BEC) \cite{Pethick2002}. Before describing the many-body phenomena in detail, it is useful to discuss the single-particle picture in order to better understand the concept of resonant driving in this model.

\subsection{Single particle problem} \label{one_particle}
Consider a classical particle bouncing on an oscillating mirror with frequency $\omega$ and amplitude proportional to $\lambda$ in the presence of a gravitational field. The Hamiltonian of the system in the frame moving with the mirror and in gravitational units has the form \cite{Sacha2015,Sacha2018,Holthaus1994,Flatte1996,Buchleitner2002}
\be
H_0(z,p,t)=\frac{p^2}{2}+z+\lambda z\cos\omega t, \quad z\ge 0.
\label{h}
\ee
For $\lambda=0$, the system is integrable and its motion is periodic with frequency $\Omega=\partial H_0(I)/\partial I$, where $H_0(I)=(3\pi I)^{2/3}/2$ is the unperturbed part of the Hamiltonian (\ref{h}) in the action-angle variables \cite{Lichtenberg1992,Flatte1996,Buchleitner2002,Sacha2018}. In this picture, $I$ is the classical analogue of the energy quantum number for an unperturbed particle. The distance of the mirror from the classical turning point is related to the frequency via the relation $h_0=\pi^2/(2\Omega^2)$.

In the presence of small mirror oscillations ($\lambda \ll 1$), we are interested in the motion of a particle sufficiently close to a periodic orbit that is resonant with the driving, i.e., $\omega=n\Omega$, where $n$ is an integer number. Because of the periodicity in time of the system, in the quantum description we can define the Floquet Hamiltonian $H=H_0-i \partial_t$ which possesses $T$-periodic eigenstates (where $T=2\pi/\omega$) known as Floquet states \cite{Flatte1996,Buchleitner2002}. To describe the motion of a particle close to the $n:1$ resonant orbit one can apply the secular approximation \cite{Berman1977,Lichtenberg1992,Flatte1996,Buchleitner2002,Sacha2018}. The resulting secular Hamiltonian indicates that in the frame moving along a resonant orbit, an atom behaves effectively like a particle in a periodic time-independent lattice potential with $n$ sites and periodic boundary conditions and for $n \to \infty$ a band structure in the quasi-energy spectrum emerges \cite{Guo2013,sacha2015anderson,Guo2016,Guo2016a,
Liang2017,Sacha2018}. We only consider the first quasi-energy band, therefore we construct $n$ Wannier functions $w_i(z,t)$ localized in different sites of the periodic effective potential \cite{sacha2015anderson,Sacha2018}. These Wannier functions, in the laboratory frame, are localized wavepackets $w_i(z,t)$ moving along a classical resonant orbit with period $nT$. Now we switch from single particle to ultra-cold atoms which fulfill an $n:1$ resonance condition with the mirror motion. 

\subsection{Cloud of ultra-cold atoms}
\label{n_particle}
The many-body Floquet Hamiltonian of ultra-cold bosonic atoms which are bouncing resonantly on an oscillating atom mirror, in the Hilbert subspace of $(nT)$-periodic states, can be written in the form \cite{Sacha2015,sacha2015anderson,Sacha2018}
\be
\hat{\cal H}=\frac{1}{nT}\int\limits_0^{nT}dt\int \limits_0^\infty  dz\;\hat\Psi^\dagger\left[H_0 +\frac{g_{0}}{2}\hat\Psi^\dagger\hat\Psi-i\partial_t\right]\hat\Psi ,
\label{H_N_body}
\ee  
where $H_0$ is the single-particle Hamiltonian (\ref{h}) and $\hat{\Psi}$ is the bosonic field operator. For a BEC all atoms occupy the same single-particle state and the many-body wave-function factorizes as $\phi_0(z_1,t)\phi_0(z_2,t) ... \phi_0(z_N,t)$. In the mean-field approximation $\phi_0(z,t)$ is a solution of the Gross–Pitaevskii (GP) equation \cite{Pethick2002}
\be
i\partial_t\phi_0(z, t)=\left[\frac{p^2}{2}+z+\lambda z\cos\omega t+g_{0}N|\phi_0(z,t)|^2\right]\phi_0(z, t).
\label{gpfull}
\ee
In order to get intuition about solutions of the GP equation that describe resonant motion of atoms let us restrict the analysis to the resonant single-particle Hilbert subspace spanned by the $n$ localized Wannier wavepackets $w_i(z,t)$ of the first quasi-energy band of the $n:1$ resonantly driven system (\ref{h}), see \cite{Sacha2015,sacha2015anderson,Sacha2018} for details. The mean-field energy functional corresponding to the many-body Floquet Hamiltonian (\ref{H_N_body}), in the Wannier basis $\phi_0(z, t)=\sum_{i=1}^n a_i w_i(z, t)$, has the form
\be
E=-\frac{J}{2}\sum_{i=1}^n\; (a_{i+1}^* a_{i}+c.c.)+\frac{N}{2}\sum_{i,j=1}^n U_{i j} |a_i|^2  |a_{j}|^2,
\label{ener1} 
\ee 
where $J=-\frac{2}{nT}\int_0^{nT}dt\int_0^\infty dz\;w^*_{i+1}(z, t) [H_0 - i \partial_t] w_{i}(z, t)$ is the tunneling amplitude of atoms between neighboring Wannier wavepackets while $U_{ij}=\frac{2g_0}{nT}\int_0^{nT}dt\int_0^\infty dz\;|w_{i}(z, t)|^2|w_{j}(z, t)|^2$ for $i\ne j$ and $U_{ii}=\frac{g_0}{nT}\int_0^{nT}dt\int_0^\infty dz\;|w_{i}(z, t)|^4$ describe the strength of the effective interactions between atoms. This energy $E$ is actually the mean-field quasi-energy per particle. Extrema of $E$ are given by solutions of the GP equation (\ref{gpfull}) and can be found analytically for the case $n=2$ \cite{Sacha2015}, and numerically for $n>2$. It turns out that if the strength of the attractive interactions is smaller than a certain critical value, i.e. $|g_0N|<|g_{cr}N|$, the mean-field solution corresponding to the minimal energy $E$ is of the form $\phi_0(z,t)=(1/\sqrt{n})\sum_{i=1}^n w_i(z,t)$ \cite{Sacha2015}. However, when the interaction strength is larger than the critical value $|g_{cr}N|$, $\phi_0(z,t)$ is not a uniform superposition of $w_i(z,t)$, which means that the system chooses a periodic solution evolving with the period $n$ times longer than the period expected from the symmetry of the Hamiltonian (see Fig.~\ref{setup} for the case $n=4$). The discrete time translation symmetry is broken and a period $n$-tupling time crystal phase forms. The solution for sufficiently large interaction is given by the single wavepacket $\phi_0(z,t) \approx w_i(z,t)$. For this reason it is much better to realize the experiment in the regime $|g_0N| \gg |g_{cr}N|$ \cite{Sacha2018}.

\subsection{Challenges in the realization of a period $n$-tupling DTC} \label{Challenges}
In this subsection we discuss the most important challenges in the realization of a period $n$-tupling DTC by means of ultra-cold atom system bouncing on the oscillating mirror. In the laboratory it could be difficult to realize the hard-wall mirror that we have assumed in all theoretical analyses. However, if a realistic Gaussian shape mirror (that can be produced by a repulsive light-sheet) is used, the same time crystal phenomena can be realized as in the hard-wall case. \cite{Giergiel2020}.

In the experimental realization of a DTC, the initial distribution of BEC will be prepared as a harmonic oscillator ground state matching the Wannier state at a classical turning point \cite{Sacha2018}. However, the preparation of this initial atomic distribution is subject to experimental imperfections. For example, the atomic cloud released from the harmonic trap may have non-zero initial momentum, can be displaced from the classical turning point and may have non-Gaussian shape. The displacement of the cloud introduces detuning of bouncing gas of atoms from the resonant driving by the oscillating mirror. Moreover, there are typically shot-to-shot fluctuations in the position of the atoms in subsequent repetitions of an experiment. We expect that a sufficiently strong attractive interaction between atoms will compensate small displacements of the initial position of the wavepacket from the classical turning point. To show the robustness of DTC against perturbations of the initial atomic distribution, we determine the phase diagram as a function of displacement and interaction strength, see Sec. \ref{ph_diag}. However, if the displacement parameter is larger than the critical value and the initial position is not stable the mirror frequency needs to be corrected. For determining the optimal mirror frequency we have used the Bayesian optimization method, see Sec. \ref{opt}.

It is also important to take into account the potential heating sources as well as possible atomic losses occurring in the system. Hence, we examine the stability of DTC on a long time scale against quantum many-body fluctuations which can result in heating of the system, see Sec. \ref{Bogoliubov}. In order to reduce atomic losses, it is better to choose a higher ratio of response period to driving period \cite{Sacha2018}. For a larger value of $n$ the numerical simulations are time consuming. Therefore we also discuss the Bayesian optimization in the context of reducing numerical cost which would be essential for potential experiments, where a higher $n$ is required \cite{Sacha2018,Giergiel2020}. 

In order to give a flavour of what experimental parameters are needed to realize the $n$-tupling DTC let us consider an example of $n=30$ \cite{Giergiel2020}. Initially, a BEC of $N\approx 5000$ $^{39}$K atoms should be prepared in a spherical harmonic trap of frequency 95~Hz at a distance of 145~$\mu\rm m$ from a horizontal blue-detuned (532~nm) light sheet. The latter produces a Gaussian shape mirror potential. The s-wave scattering length $a_s$ of $^{39}$K $|1,+1\rangle$ atoms is controlled by means of the Feshbach resonance with the centre at a magnetic field of 402.5~G and the zero crossing point at 350.5~G. When the trapping potential along the vertical (longitudinal) direction is turned off, the width of the initial atomic cloud for $a_s\approx 0$ matches the width of the Wannier wavepacket at the classical turning point. The confinement along the transverse directions is kept during the entire experiment. If the width of the Gaussian mirror potential is $10~\mu\rm m$ and its height $3.7\times 10^3$ in the gravitational units, the amplitude and frequency of the mirror oscillations suitable for the realization of the DTC are 75~nm and 2.8~kHz, respectively. If the atomic s-wave scattering length $a_s\approx 0$, atoms bouncing on the oscillating mirror will tunnel from the initial Wannier wavepacket to the neighbouring wavepackets after about 50 bounces. However, the same experiment performed for $a_s=-1.6a_0$ where $a_0$ is the Bohr radius, will show stable evolution of the atomic cloud with a period 30 times longer than the mirror oscillation period, demonstrating a $30$-tupling DTC. The 50 bounces of atoms on the mirror last about 0.6~s. The estimated lifetime of the atomic cloud due to three-body recombination is $\tau_3=[\langle\rho^2\rangle K_3]^{-1}\approx 10(4)$~s for the chosen parameters where $\langle\rho^2\rangle\approx 8\times 10^{27}$~cm$^{-6}$ and $K_3=1.3(5)\times 10^{-29}$~cm$^6$~s$^{-1}$ \cite{Fattori} near the zero crossing point of the s-wave scattering length. The lifetime $\tau_3$ is sufficiently long to demonstrate the DTC. Losses due to collisions with background atoms can be kept negligible if a high quality vacuum is used. Losses due to photon scattering for light from the far detuned 532~nm light sheet mirror should be negligible.

\section{Constructing the phase diagram for period-doubling and period-quadrupling discrete time crystals} \label{ph_diag}
We investigate the robustness of DTCs with regards to imperfect preparation of the initial state of the BEC. We start from an initial state which is the optimal Gaussian approximation of the mean-field solution of the DTC located exactly at the classical turning point $h_0$ above the mirror
\be
\phi_0(z,t=0)=\left(\frac{\tilde\omega_0}{\pi}\right)^{1/4}e^{-\tilde\omega_0(z-h_0)^2/2}.
\label{w_p}
\ee
The parameter $\tilde\omega_0$ is the frequency of the harmonic trap where the BEC is initially prepared and it is chosen such that the width of the atomic cloud matches the width of the Wannier wavepacket $w_i(z, t)$ at the classical turning point \cite{Sacha2018}. We consider an $n:1$ resonant driving of atoms by an oscillating atom mirror (see Sec. \ref{one_particle}). Spontaneous breaking of the discrete time translation symmetry of the system, and consequent formation of the DTC, occurs for sufficiently strong attractive boson-boson interactions \cite{Sacha2015,Sacha2018}. Then, the initially prepared wavepacket (\ref{w_p}) is expected to be returning to the vicinity of the initial position after each period $nT$, as one can observe for the case $n=4$ in Fig.~\ref{setup}.
\begin{figure}[ht!] 	    
\centering        
\includegraphics[]{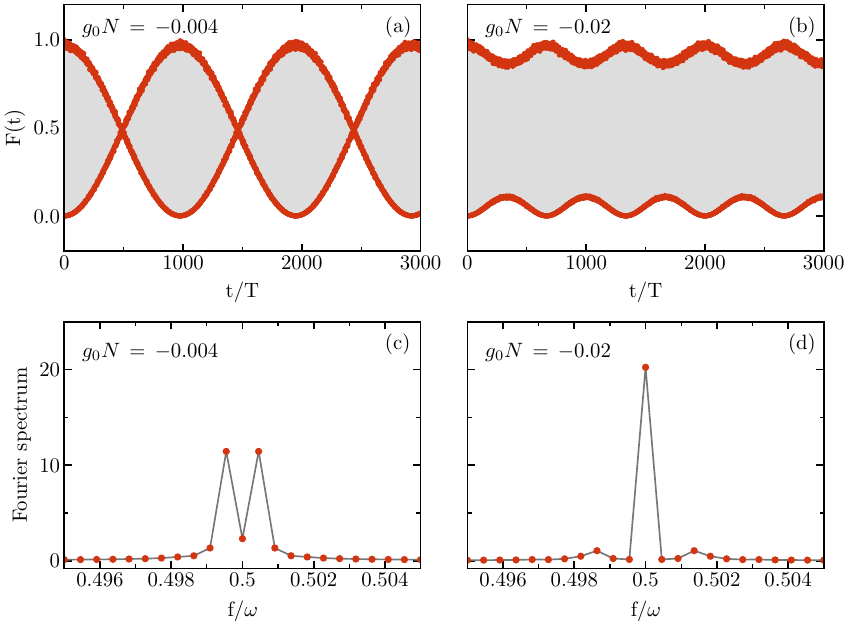}    
\caption{Period-doubling DTC ($n=2$). Top panels: Quantum fidelity (\ref{fid}) obtained for the initial state (\ref{w_p}) as a function of $t/T$ for $g_0 N=-0.004$ (a) and $g_0 N=-0.02$ (b). In (a), the evolution of the fidelity reveals beating which is related to tunneling of atoms between two Wannier wavepackets $w_{1,2}(z,t)$ that evolve along the $2:1$ resonant orbit. The beating period corresponds to the tunneling period $\pi/J$ where $J=7.26 \times 10^{-4}$. Bottom panels: Fourier transform of the fidelity obtained at the stroboscopic time $t=T,2T,3T...$ for $g_0 N=-0.01$ (c) and $g_0 N=-0.02$ (d). One can observe the peaks of the Fourier transform located at $f=\omega/2$ for $|g_0N| \gg |g_{cr} N|$, where $g_{cr}N \approx -0.006$. The parameters of the initial state (\ref{w_p}) are: $\tilde\omega_0=0.68$ and $h_0=9.82$ and the frequency and the amplitude of the mirror oscillations are $\omega=1.4$ and $\lambda=0.12$, respectively.}
\label{fidelity_FT}   
\end{figure}

It is convenient to introduce the quantum fidelity function
\be
F(t)=|\langle\phi_0(0)|\phi_0(t)\rangle|^2,
\label{fid}
\ee
where $\phi_0(t)$ is the solution of the GP equation (\ref{gpfull}). In an experiment such a quantity can be recovered from the measurement of the atomic density $\rho(z,t)=N|\phi_0(z,t)|^2$. Indeed, $F(t)\propto \int dz \rho(z,0) \rho(z,t) $ because $\rho(z,t)$ is always a localized distribution. The fidelity (\ref{fid}) for $n=2$ (thus for the period-doubling DTC) and for sufficiently strong attraction is presented in Fig.~\ref{fidelity_FT}b. The Fourier transform of $F(t)$ shows a single peak located at the half driving frequency $\omega/2$ (Fig.~\ref{fidelity_FT}d). This peak is related to the subharmonic response of the system which is the signature of the period-doubling time crystal \cite{Yao2017}. However, for weak attractive interactions (which correspond to the symmetry preserving regime) after the tunneling time $t=\pi/J\gg T$ we observe transfer of atoms to the second wavepacket (Fig.~\ref{fidelity_FT}a) which evolves along the same $2:1$ resonant trajectory and is delayed (or advanced, depending on the point of view) by $T$ with respect to the initial wavepacket. After another period $\pi/J$, atoms tunnel back to the initial wavepacket and this dynamics continues with period $2\pi/J$. Consequently, the Fourier transform of the fidelity function (\ref{fid}) reveals splitting of the Fourier peak around $\omega/2$ (Fig.~\ref{fidelity_FT}c). Such two separated peaks in the frequency domain are consistent with the beating in the plot of $F(t)$. Note that the decay of the $(2T)$-periodic evolution due to tunneling of non-interacting atoms takes place even if there is no detuning from the resonant driving. It is in contrast to DTCs in spin systems where in the absence of any detuning from the perfect spin flip, time evolution of the spin systems is still periodic with period $2T$ \cite{PhysRevLett.116.250401,PhysRevLett.117.090402,Yao2017,Monroe2017,choi2017observation,Pal2018,Rovny2018}.

As mentioned in Sec. \ref{Challenges} various experimental imperfections may cause fluctuations in the average momentum of atoms and of the frequency of the harmonic trap where the atomic cloud is initially prepared. Hence, we consider the initial states
\be
\phi_0(z,t=0)=\left(\frac{\tilde\omega}{\pi}\right)^{1/4}e^{-\tilde\omega(z-h)^2/2- i p_0(z-h)},
\label{w_p2}
\ee
where the initial average momentum $p_0$ and the frequency $\tilde\omega$ are sampled from the uniform distributions in the intervals $(-\delta p_0, \delta p_0)$ and $(\tilde\omega_0-\delta \tilde\omega_0,\tilde\omega_0+ \delta \tilde\omega_0)$, respectively. Furthermore the initial location of the wavepacket (\ref{w_p2}) is taken to be $h = h_0 + \epsilon$, where the displacement parameter $\epsilon$ denotes the detuning of the system from a perfect resonant driving. This displacement parameter $\epsilon$ and the interaction strength $g_0N$ constitute the space of parameters for which we determine the phase diagram of the DTC. We expect that even if $\epsilon\ne 0$, sufficiently strong interactions are able to stabilize the evolution of the DTC.

\begin{figure} [ht!]      
\centering      
\includegraphics[]{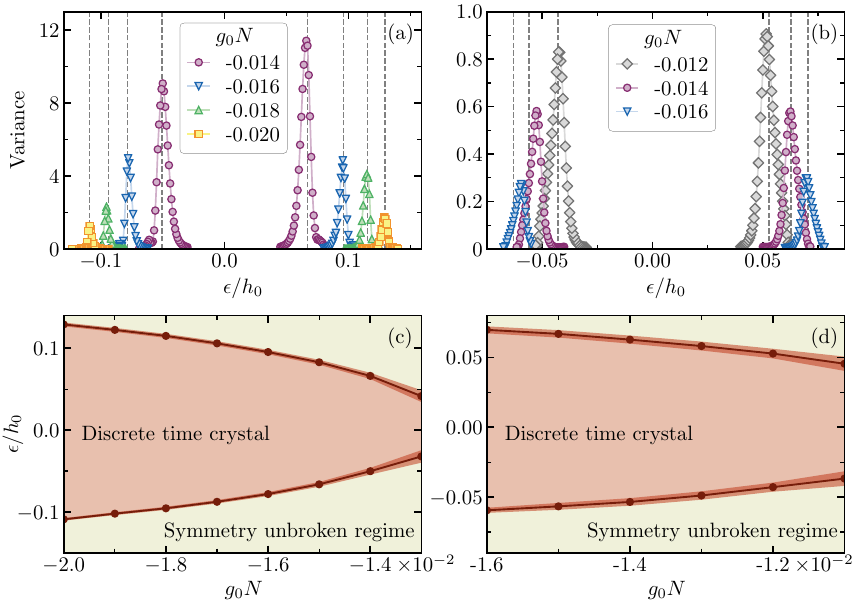}           
\caption{Period-doubling ($n=2$) and period-quadrupling ($n=4$) DTCs. Top panels: Variance of the Fourier peak magnitude as a function of displacement parameter $\epsilon$ for different values of $g_0N$ (given in the legend) with $n=2$ (a) and $n=4$ (b). Bottom panels: Phase diagram of DTC in the space of the displacement parameter $\epsilon$ and interacting strength $g_0N$ for $n=2$ (c) and $n=4$ (d). Orange shaded regions around the phase boundary solid lines show the standard deviation of the variance around its maximal values, cf. top panels. In the $n=2$ case, the parameters are the following: $\omega=1.4$, $\lambda=0.12$, $h_0=9.82$, $\tilde\omega_0=0.68$ while $p_0$ and $\tilde\omega/\tilde\omega_0$ in (\ref{w_p2}) are uniformly drawn in the intervals $[-0.1,0.1]$ and $[0.98,1.02]$, respectively. In the $n=4$ case, we have: $\omega=1.79$, $\lambda=0.12$, $h_0=24.44$, $\tilde\omega_0=0.5$, $p_0\in[-0.05,0.05]$ and $\tilde\omega/\tilde\omega_0\in[0.98,1.02]$. }
\label{var_PD}   
\end{figure} 

To determine the critical value of $\epsilon$ for a given $g_0N$, we analyze the fluctuations of the amplitude of the peak at $\omega/2$ (for period-doubling DTC) and at $\omega/4$ (for period-quadrupling DTC) in the Fourier transform of the fidelity (\ref{fid}) obtained in $m$ random realizations of (\ref{w_p2}). We expect that the largest fluctuations of the peak amplitude can be observed near the critical point between the DTC regime and the symmetry unbroken regime \cite{Yao2017}.  Indeed, the variance of the peak amplitude shows a strong maximum at the critical point as clearly visible in Fig.~\ref{var_PD}a-b. Performing numerical simulations for different values of $g_0N$ we obtain the phase diagram depicted in Fig.~\ref{var_PD}c for the period-doubling DTC. Similar phase diagram but for the period 4-tupling DTC is shown in Fig.~\ref{var_PD}d. In the latter case the maximum of the variance of the amplitude of the Fourier peak at $\omega/4$ is used as the signature of the critical point. One can see that the phase diagrams are not symmetric with respect to $\epsilon =0$. It shows that the displacement with the case $h>h_0$ is favorable for the formation of a DTC, compared to the case $h<h_0$. This is a consequence of the influence of the gravitational field. It should be noticed that the above approach is used in the regime where the symmetry breaking state is approximately given by a single wavepacket. Close to $g_{cr}N$, which corresponds to the perfect $\epsilon=0$ case, the symmetry breaking states are a superposition of $n$ wavepackets with unequal weights. Consequently, detuning of the system from the perfect resonant driving cannot be described by $\epsilon$ only. Indeed, one should also consider the relative displacement of the wavepackets and deviation from the optimal relative phase between them. Therefore, in order to obtain $g_{cr}N$, we have used the $n$-mode approximation (\ref{ener1}).

To reduce the numerical burden, locating the critical $\epsilon$, corresponding to a maximum of the variance, was turned into an optimization problem for which we have applied Bayesian optimization (see \ref{BO}). The positions of the variance peaks predicted at the end of the optimizations are marked by the dashed vertical lines in Fig. \ref{var_PD}a-b. Using Bayesian optimization for 15 iterations to obtain the phase diagrams in Fig. \ref{var_PD}c-d we have reduced by tenfold the computational time with efficiency around 70\%.

We have presented the phase diagrams for $n=2$ and $n=4$. In order to obtain the phase diagrams for other values of $n$, numerical simulations are required. However, one can estimate how the width of the diagrams scales with $\lambda$ and $n$ by analyzing the size of the classical resonant islands of the single-particle system and assuming that each of them is large enough to support two quantum states \cite{Sacha2018, Buchleitner2002}. Such an analysis allows us to determine the range of displacement from the classical turning point $\Delta \epsilon$ which ensures that particle motion remains in the vicinity of the resonant orbit. The result, in the leading order of $\lambda$, is the following
\be
\frac{\Delta\epsilon}{h_0} \approx \frac{8\sqrt{\lambda}}{n \pi},
\ee
with the constraint $\lambda \le 0.2$ which ensures that classical motion around the resonant orbit is not chaotic.

\section{Optimal control of the distance of the atomic cloud to the mirror} \label{opt}
In an experimental realization of a DTC, the true distance $h$ to the mirror may deviate more than the critical $\epsilon$ from $h_0$ (see Sec. \ref{ph_diag}) and will fluctuate around an average value $\bar h$ in between each realization. Here we show that the mirror oscillations frequency could be adjusted, directly onto the experiment, to best account for this unknown average value~$\bar h$.

For that purpose, we resort to Bayesian optimization in order to optimize the mirror frequency which allows one to control the true distance $h$. In the following we assume that the atomic cloud is prepared at a distance $h$ randomly chosen, at each realization, from the uniform distribution in the range $[\bar h-\delta h,\bar h+\delta h]$. Because we do not know $\bar h$ we choose the frequency $\tilde\omega_0$ of the harmonic trap optimal for $h_0$. Furthermore, $\bar h$ is taken to be such that $\bar h-h_0\approx 0.3h_0\gg \delta h$. In practice it is important to ensure that $\bar h$ is above $h_0$ because if $ \bar{h}-h_0 < -0.1 h_0$, there is significant overlap of $\phi_0(z, 0)$ with more than one Wannier state limiting the optimization procedure. To define the optimization problem one needs to specify a figure of merit to be maximized. In this case it is taken to be the fidelity function
\be
F(\omega)=|\la\phi_0(0)|\phi_0(nT)\ra|^2, 
\ee
obtained after $n$ periods of the mirror oscillations $T=2\pi/\omega$, where $\phi_0(z,t=0)$ is given in (\ref{w_p2}). At each step of the optimization this fidelity is averaged over $20$ repetitions, and this average is fed into the optimizer. At the end of the optimization an optimal frequency $\omega_{opt}$ is returned by the optimizer (see \ref{BO}).

\begin{figure} [ht!]      
\centering      
\includegraphics[]{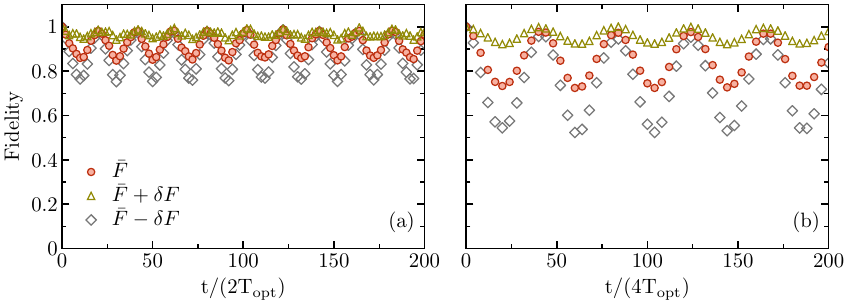}    
\caption{Figure shows fidelity for the optimized frequency $\bar F=\langle F_j(\omega_{opt}) \rangle$ averaged over 20 random realizations of the experiment together with $\bar F\pm \delta F$, where $\delta F$ is the standard deviation, at discrete moments of time where $T_{opt}=2\pi/\omega_{opt}$ --- full circles correspond to $\bar F$, open triangles to $\bar F+\delta F$ and open squares to $\bar F-\delta F$. Fifteen iterations of Bayesian optimization are used to obtain optimal $\omega_{opt}$. The left panel corresponds to the period-doubling ($n=2$) and the right panel to period-quadrupling ($n=4$) DTCs. The other parameters are the following: $\bar h=12.78$ ($31.77$), $\delta h=0.05\bar h$ ($0.03\bar h$), $\delta\tilde\omega_0=0.02\tilde\omega_0$ ($0.02\tilde\omega_0$), $\delta p_0=0.1$ ($0.05$) and $\lambda=0.12$ (0.12) in the left (right) panel.}
\label{s2_opt}   
\end{figure}

In Fig.~\ref{s2_opt}, the average values of the fidelity are plotted as a function of discrete time $t=k(nT_{opt})$, where $k$ is an integer number, $T_{opt}=2\pi/\omega_{opt}$ and $\omega_{opt}$ is the optimized frequency obtained after 15 iterations. It can be observed that the average fidelity is above 70\% for a very long time for both the period-doubling ($n=2$) and period-quadrupling ($n=4$) cases. Without optimization the fidelity drops to almost zero after a few periods of the mirror oscillations. This shows the effectiveness of the Bayesian optimization technique based on a limited number of experimental repetitions to control the true distance $h$ of the atomic cloud to the mirror.

\section{Quantum many-body fluctuations of discrete time crystals}
\label{Bogoliubov}
So far we have performed the analysis of the system within the mean-field approximation, i.e. according to the GP equation (\ref{gpfull}). These results allowed us to obtain the phase diagrams which determine how strongly one may perturb the system and still DTCs are stable. However, the mean-field approach assumes that time evolution of a Bose system is restricted to the many-body Hilbert subspace of product states, $\phi_0(z_1,t)\phi_0(z_2,t)\dots\phi_0(z_N,t)$, which correspond to a BEC. Interactions between bosons couple the product state subspace to the complementary space and can lead to decay of a DTC. We address this problem in the present section within the Bogoliubov approach \cite{Pethick2002} in the case of the period-doubling time crystal. That is, we investigate if a BEC loaded to the $2:1$ resonant orbit and evolving along such an orbit can suffer from quantum many-body fluctuations and eventually heats up.

Before we switch to the Bogoliubov description, let us first comment on the behavior of ultra-cold atoms bouncing on an oscillating mirror, when we restrict to the many-body resonant Hilbert subspace \cite{Sacha2015,Sacha2018}. In the case of the $2:1$ resonant driving, the resonant Hilbert subspace is spanned by the two Wannier-like wavepackets $w_{1,2}(z,t)$, i.e. the bosonic field operator in (\ref{H_N_body}) is restricted to $\hat\Psi(z,t)\approx w_1(z,t)\hat a_1+w_2(z,t)\hat a_2$ where $\hat a_{1,2}$ are the standard bosonic annihilation operators. The many-body Floquet Hamiltonian (\ref{H_N_body}) in the resonant subspace reads \cite{Sacha2015}
\be
\hat{\cal H}\approx-\frac{J}{2}\left(\hat a_1^\dagger\hat a_2+\hat a_2^\dagger\hat a_1\right)+\frac{U_{11}-U_{12}}{2}\left(\hat a_1^\dagger\hat a_1^\dagger\hat a_1\hat a_1+\hat a_2^\dagger\hat a_2^\dagger\hat a_2\hat a_2\right),
\label{h_two_mode}
\ee
with $J$ and $U_{ij}$ similar to (\ref{ener1}). The above Hamiltonian $\hat{\cal H}$ is identical to the Hamiltonian for bosons in a double well potential within the two-mode approximation \cite{Milburn1997} --- in the present case the two modes are not time-independent functions but the ($2T$)-periodic Wannier wavepackets $w_{1,2}(z,t)$. It is known that the Hamiltonian \eqref{h_two_mode} can be mapped to the Lipkin-Meshkov-Glick model \cite{Ribeiro2008}
\be
\hat {\cal H}=J\left(-\hat S_x+\frac{\gamma}{N}\hat S_z^2\right),
\label{lmg}
\ee 
with a constant term omitted, where the spin operators read 
\be
\hat S_x=\frac{\hat a_1^\dagger\hat a_2+\hat a_2^\dagger\hat a_1}{2}, \quad \quad
\hat S_z=\frac{\hat a_2^\dagger\hat a_2-\hat a_1^\dagger\hat a_1}{2},
\ee
and $\gamma=N(U_{11}-U_{12})/J\propto g_0N$. In the limit when $N\rightarrow\infty$ but $\gamma=\rm const$ and $\gamma<-1$ (i.e. for sufficiently strong attractive interactions between bosons), the Lipkin-Meshkov-Glick model reveals a quantum phase transition where the $\mathbb{Z}_2$ symmetry (the $\mathbb{Z}_2$ symmetry means that $\hat {\cal H}$ commutes with $e^{i\pi\hat S_x}$) of the Hamiltonian (\ref{lmg}) is spontaneously broken. Then, all the eigenstates of the model for eigenenergies below the so-called symmetry broken edge $-JN/2$ reveal spontaneous breaking of the $\mathbb{Z}_2$ symmetry. The spontaneous breaking of the $\mathbb{Z}_2$ symmetry of the model corresponds to the spontaneous breaking of the time translation symmetry of ultra-cold atoms bouncing on an oscillating mirror which, in the basis of the ($2T$)-periodic Wannier functions $w_{1,2}(z, t)$, are described by the Floquet Hamiltonian (\ref{h_two_mode}) \cite{Sacha2015}. Thus, all Floquet many-body states of the Floquet Hamiltonian (\ref{h_two_mode}) with quasi-energies below the symmetry broken edge reveal period-doubling time crystal behaviour. 

We have mentioned that the Hamiltonian of the form of (\ref{h_two_mode}) describes also bosons trapped in a symmetric double well potential \cite{Milburn1997}. The spontaneous breaking of the time translation symmetry we consider here corresponds to the self-trapping phenomenon in the double well problem. That is, if attractive interactions between bosons are sufficiently strong, they prefer to group in one of the potential wells. The symmetric ground state of the system is a macroscopic superposition of all bosons in the first potential well and all bosons in the other well, i.e. $|\psi_+\ra\approx(|N,0\ra+|0,N\ra)/\sqrt{2}$ for $\gamma\ll -1$, which is a Schr\"odinger cat-like state. Experimental realization of the state $|\psi_+\ra$ is not attainable because it is sufficient, e.g., to measure the position of one boson and the symmetric ground state collapses to $|N,0\ra$ or $|0,N\ra$ depending on which well the boson is detected  --- this is the self-trapping phenomenon which breaks the $\mathbb{Z}_2$ symmetry of the double well problem. The lifetime of the symmetry broken state, $|N,0\ra$ or $|0,N\ra$, is determined by the energy splitting between the ground state energy $E_+$ and the energy $E_-$ of the eigenstate $|\psi_-\ra\approx(|N,0\ra-|0,N\ra)/\sqrt{2}$. In the limit of $N\rightarrow\infty$ but $\gamma=\rm constant$, the lifetime behaves like $(E_+-E_-)^{-1}\propto e^{\alpha N}/N$ where $\alpha$ is a positive constant \cite{Sacha2015}. Thus, within the two-mode model, i.e. within the resonant Hilbert subspace, we obtain that the lifetime of the DTC increases nearly exponentially quickly with $N$ and in the limit of $N\rightarrow\infty$ the DTC lives forever.

Even if in the resonant many-body Hilbert subspace the formation of the DTC is clear, there is still a question what happens in the full many-body Hilbert space of the system, i.e. when we take into account that interactions between bosons couple the resonant subspace with the complementary space? It can be addressed by applying the Bogoliubov approach. We use the particle-number-conserving version of the Bogoliubov theory \cite{Castin1998} where the bosonic field operator is decomposed into the operator $\hat a_0$, corresponding to the condensate mode $\phi_0(z, t)$, and the operator $\delta\Psi_\perp(z, t)$ living in the orthogonal subspace, i.e. $\hat\Psi(z, t)=\phi_0(z, t)\hat a_0+\delta\hat\Psi_\perp(z, t)$. The operator $\delta\hat\Psi_\perp(z, t)$ describes quantum many-body fluctuations around a many-body product state. Within the mean-field approximation the DTC is a BEC evolving with the period $2T$ along the $2:1$ resonant orbit. Within the Bogoliubov approach we also start with a BEC, i.e. we assume that  at $t=0$ all bosons are in a many-body product state $\Phi(z, 0)=\phi_0(z_1,0)\phi_0(z_2, 0)\dots\phi_0(z_N, 0)$, and we expect that in the course of time evolution of atoms along the resonant orbit the interactions between bosons can lead to quantum depletion of a BEC. Initially we have a perfect condensate and consequently eigenvalues of the reduced single-particle density matrix, $\rho(z,z';t=0)=\la\Phi(0)|\hat\Psi^\dagger(z,t=0)\hat\Psi(z',t=0)|\Phi(0)\ra$, are all zero except the one corresponding to a condensate mode $\phi_0(z, t)$ which is equal to the total number of particles $N$. In the course of time evolution, bosons are being depleted from a condensate mode $\phi_0(z, t)$ and start occupying other modes which is indicated by the fact that not only one eigenvalue of the reduced single-particle density matrix is non-zero. In the Bogoliubov approach the total number of bosons $dN(t)$ depleted from the condensate is equal to the sum of the norms of $v_i(z,t)$ components of Bogoliubov modes $[u_i(z,t),v_i(z,t)]$,
\be
dN(t)=\sum_i\la v_i(t)|v_i(t)\ra.
\ee
In the particle-number-conserving version of the Bogoliubov theory \cite{Castin1998}, the modes evolve according to the following linear Bogoliubov-de~Gennes equation
\be 
i\partial_t\left[\begin{matrix} u_i \\ v_i \end{matrix}\right]=
\left[\begin{matrix}
\hat Q\left(H_0+2g_0N|\phi_0(z,t)|^2\right)\hat Q & g_0N\hat Q\phi_0^2(z,t)\hat Q^* \\
-g_0N\hat Q^*{\phi_0^*}^2(z,t)\hat Q&-\hat Q^*\left( H_0+2g_0N|\phi_0(z,t)|^2\right)\hat Q^*
\end{matrix}\right]\left[\begin{matrix} u_i \\ v_i \end{matrix}\right],
\label{BdG}
\ee
where $\phi_0(z,t)$ fulfills the GP equation (\ref{gpfull}), $\hat Q=1-|\phi_0(t)\ra\la\phi_0(t)|$ and $\hat Q^*=1-|\phi_0^*(t)\ra\la\phi_0^*(t)|$.
Initial Bogoliubov modes allow us to define an initial many-body state of $N$ bosons. If at $t=0$ we choose $[u_i(z,0),v_i(z,0)]=[\chi_i(z),0]$ (where $\la\chi_i|\chi_j\ra=\delta_{ij}$ and  $\la \chi_i|\phi_0(0)\ra=0$) and the Bogoliubov vacuum state as the initial state \cite{Dziarmaga2006}, we deal with the many-body state which is a perfect condensate \cite{Dziarmaga2006}, i.e. no bosons are initially depleted $dN(0)=0$. The choice of $\chi_i(z)$ is arbitrary but if we do our best and choose $\chi_i(z)$ optimally adapted to a given problem, we will have to evolve a small number of the Bogliubov modes only in order to get the converged result for the total number of depleted bosons $dN(t)$. In the case of the $2:1$ resonant bouncing of ultra-cold atoms on an oscillating atom mirror, we start with $\phi_0(z,0)$ as the Gaussian state (\ref{w_p}) and $\chi_i(z)$ as eigenstates of the Hartree-Fock Hamiltonian $H_{\text{HF}}=-\frac12\partial^2_z+z+\lambda z+2g_0N|\phi_0(z,0)|^2$ where contributions to the condensate mode are subtracted, i.e. $\la \chi_i|\phi_0(0)\ra=0$ and $\chi_i(z)$  are corrected so that $\la\chi_i|\chi_j\ra=\delta_{ij}$. 

\begin{figure} [ht!]      
\centering      
\includegraphics[]{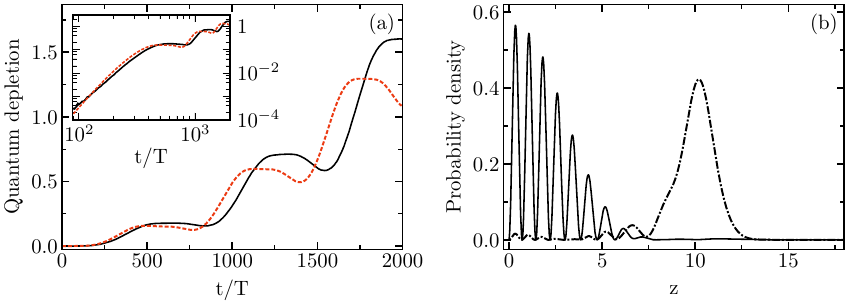}           

\caption{Left panel: black solid line shows the average number of atoms $dN(t)$ depleted from the condensate wavefunction $\phi_0(z, t)$. Initially $\phi_0(z, t)$ corresponds to the Gaussian wavepacket (\ref{w_p}) and its time evolution $\phi_0(z, t)$, according to the GP equation (\ref{gpfull}), describes the period-doubling time crystal. The total depletion of the condensate is dominated by a single eigenmode $\phi_1(z, t)$ of the reduced single-particle density matrix (\ref{rho1}), i.e. $dN(t)\approx dN_1(t)$, because other eigenvalues $dN_{j>1}(t)\le6\times 10^{-3}$. Red dashed line shows the condensate depletion obtained with the help of the two-mode approach (\ref{h_two_mode}). Inset presents the total depletion $dN(t)$ in the log-log-scale indicating the initial algebraic increase of the depletion. Right panel: probability densities of the condensate mode $|\phi_0(z,t)|^2$ (black solid line) and the dominant mode $|\phi_1(z,t)|^2$ (dotted-dashed line) at the final moment of the time evolution, i.e. at $t=1999T$ --- both $\phi_0(z,t)$ and $\phi_1(z,t)$ are normalized to unity.
The parameters of the mirror oscillations are: $\lambda=0.12$ and $\omega=1.4$. The total evolution time $1999T$ corresponds to two tunneling periods of non-interacting atoms between the two Wannier wavepackets $w_{1,2}$ which is $2\pi/J$ where $J=7.26\times 10^{-4}$, cf. (\ref{h_two_mode}). The parameters of the two-mode Hamiltonian (\ref{h_two_mode}) are $U/g_0=0.23$ and $U_{12}/g_0=0.10$, and the interaction strength is chosen so that $g_0N=-0.02$. The two-mode results are obtained for $N=600$ but they remain the same if $N$ is greater. }
\label{Bogol1}   
\end{figure}

Evolving the condensate mode $\phi_0(z,t)$ according to the GP equation (\ref{gpfull}) and the Bogoliubov modes $[u_i(z,t),v_i(z,t)]$ according to the Bogolibov-de~Gennes equations (\ref{BdG}) we can obtain the reduced single-particle density matrix at any time $t$ and diagonalize~it,
\bea
\rho(z,z';t)&\approx & N\phi_0^*(z,t)\phi_0(z',t)+\sum_i v_i(z,t)v_i^*(z',t) \cr
&=& N\phi_0^*(z,t)\phi_0(z',t)+\sum_{j=1}^\infty dN_j(t)\;\phi_j^*(z,t)\phi_j(z',t).
\label{rho1}
\eea
Figure~\ref{Bogol1}a shows how the total depletion changes in time. It turns out that it is entirely determined by only one eigenmode of the single-particle density matrix, i.e. $dN(t)=\sum_jdN_j(t)\approx dN_1(t)$. After a long time of 1000 bounces of ultra-cold atoms on an oscillating mirror, i.e. at $t_f=1999T$, the total depletion is of the order of one atom only, $dN(t_f)\approx 1.6$. It depends on the product $g_0N$ because the Bogoliubov-de~Gennes equations (\ref{BdG}) depend on $g_0N$. In the experiment a typical total number of atoms is of the order $N\approx 10^4$ which implies that for $g_0N=-0.02$ chosen here, $dN(t_f)/N$ is negligible and the DTC is stable and resistant to quantum many-body fluctuations at least in a time scale that we have investigated here. The latter is much longer than expected duration of the experiment \cite{Sacha2018}. 

Figure~\ref{Bogol1}b presents probability densities of the condensate mode $|\phi_0(z,t_f)|^2$ and the dominant mode $|\phi_1(z,t_f)|^2$ of the reduced single-particle density matrix \eqref{rho1}. It turns out that $|\la \phi_0(t_f\pm T)|\phi_1(t_f)\ra|^2\approx 0.87\pm 0.02$ what implies that atoms depleted from the condensate occupy the wavepacket that travels along the $2:1$ resonant orbit but is delayed with respect to the condensate mode by the period $T$ of the mirror oscillations. Thus, the many-body evolution of the system is restricted to two modes and these modes are similar to the modes $w_1(z, t)$ and $w_2(z, t)$ used to define the resonant many-body Hilbert subspace, cf. (\ref{h_two_mode}). 

Let us compare the quantum many-body effects in the time crystal evolution obtained within the Bogoliubov approach and by means of the two-mode Hamiltonian (\ref{h_two_mode}). In the two-mode case, the perfect BEC with all atoms occupying the mode $w_1(z, t)$ corresponds to $|N,0\ra$ and it is the initial state we choose in the two-mode description. At $t=0$ the quantum depletion of the condensate is zero but because the initial state is not an eigenstate of the Hamiltonian (\ref{h_two_mode}), the depletion increases in time which is shown in Fig.~\ref{Bogol1}a. Despite the fact that in the Bogoliubov description, the initial condensate wavefunction $\phi_0(z,0)$ is not exactly the mode $w_1(z, 0)$ (i.e. $\phi_0(z,0)$ is the Gaussian approximation of the mode $w_1(z, t)$ only), the results for the quantum depletion obtained with the help of the both methods follow each other quite well. The results of the two-mode approach correspond to $N=600$ but they are the same for any $N>600$ provided $g_0N=-0.02$. The two-mode description allows us also to investigate what happens in an extremely long time scale. It turns out that at $t\approx 500T\sqrt{N}$, the depletion saturates at $dN\approx 0.02N$ and next, for much longer time evolution, the $N$-body system shows a revival and returns nearly perfectly to the initial BEC.

The initial BEC states used in the Bogoliubov description and in the two-mode approach are generic uncorrelated states of the resonantly driven many-body system. The presented results of the many-body time evolution of these states in the presence of the time-periodic driving show that heating effects are negligible. 

One may ask if a general $n$-tupling DTC is also resistant to heating. The $n$-mode model for the $n$-tupling DTC can be derived, see the energy functional (\ref{ener1}), and the scaling of its parameters with the parameters of the mirror oscillations can be analyzed, see Ref.~\cite{Giergiel2020} for details. Within the $n$-mode model, spontaneous formation of a DTC corresponds to self-trapping of attractively interacting bosons (for sufficiently strong attractive interaction) in a single well of the $n$-well potential and no heating of the system occurs. In order to prove that the predictions of the model for $n\ge 3$ are valid, numerical simulations, e.g., within the Bogoliubov approach are required or experimental evidence is needed. The experiments for $n=20-100$ are in progress \cite{Giergiel2020}.

\section{Conclusion} 
The intriguing possibility of studying novel quantum many-body phenomena in the time domain relies on realizing $n$-tupling DTC with large $n$, which so far has eluded any experimental observations. Adopting a model of ultra-cold atoms bouncing on an oscillating mirror that can be experimentally realized, we demonstrate the robustness of the DTC against perturbations to the initial distribution of atoms. The Bayesian optimization used here to obtain the phase diagram efficiently for small $n$ can be naturally extended for large $n$-tupling DTC. This provides invaluable information for the experimentalists as it provides a well defined criterion for distinguishing the symmetry broken phase from the unbroken phase. The optimization also allows us to control and manipulate the system despite the experimental uncertainties and imperfections. Finally, our analysis of quantum many-body fluctuations that go beyond the mean-field approximation, clearly indicates that DTCs realized in our model do not show any signs of heating on long time scale --- much longer than the duration of experiments. Thus, the integrability of the periodically driven single-particle system that reveals non-linear resonances is inherited by the many-body counterpart if the interactions between particles are weak but still sufficiently strong to form DTCs. In summary, this work is definitely a step towards bridging the gap between theory and experiments on $n$-tupling DTC with the possibility of using optimal control in the experimental realisation of DTCs in ultra-cold systems.

\section*{Acknowledgements}
We are grateful to Bryan Dalton, Peter Hannaford and Jia Wang for the fruitful discussion and valuable comments. Support of the National Science Centre, Poland via Projects QuantERA programme No.~2017/25/Z/ST2/03027 (A.K.), No.~2016/20/W/ST4/00314 and No.~2019/32/T/ST2/00413 (K.G.) and No.~2018/31/B/ST2/00349 (W.G. and K.S.) is acknowledged.

\appendix 

\section{Bayesian optimisation} \label{BO}
A typical optimization problem involves maximizing a figure of merit $F(\mathbf{x})$ with respect to its input parameters $\mathbf{x}$,
\begin{equation} \label{optim}
\mathbf{x}^{\rm opt} = \argmax_{\mathbf{x}}~F(\mathbf{x}) .
\end{equation}
In general, $\mathbf{x}$ can be a $\cal N$-dimensional vector where $\cal N$ is the total number of parameters. In order to obtain the optimum solution $\mathbf{x}^{\rm opt}$, one has to evaluate the figure of merit multiple times with $F(\mathbf{x}_i)$ representing the result of i$^{\rm th}$ evaluation. The dependence of the figure of merit $F(\mathbf{x})$ on $\mathbf{x}$ defines the optimization landscape which can often be non trivial. One way to perform this optimization task is by means of gradient methods. However, as it is the case here, analytical gradient of $F(\mathbf{x})$ are not available, and approximations of these gradients by finite differences require extra numerical or experimental effort. Moreover this approach is limited by the presence of local extremum in the optimization landscape. 

Keeping this in mind, we resort to a non-gradient based optimization scheme, namely Bayesian optimization, which has been extremely efficient in terms of the number of evaluations needed to obtain close-to-optimum solutions \cite{wigley2016fast, zhu2018training, Henson2018, Nakamura2019} even in the presence of noise in the input parameters \cite{mukherjee2020preparation}, or in the evaluation of the figure of merit \cite{sauvage2019}. Bayesian optimization is a technique that adopts a probabilistic approach towards optimization. It has essentially three steps: first, it approximates the unknown optimization landscape $F(\mathbf{x})$ with well-behaved random functions $f$ with prior distribution, that is before obtaining any evaluation, $p(f)$. Given $M$ evaluations of the figure of merit denoted $\mathbf{y}_M=[F(\mathbf{x}_1),\hdots F(\mathbf{x}_M)]$, this distribution is updated to incorporate the data acquired by means of Bayes' rule
\begin{equation} \label{update}
p(f|\mathbf{y}_{M}) = \frac{p(f) p(\mathbf{y}_{M}|f)}{p(\mathbf{y}_{M})} ,
\end{equation}  

The final step involves deciding, based on this model $f$, which parameters to evaluate next. A short-sighted strategy would consist on selecting the next set of parameters $\mathbf{x}_{i+1}$ where the model takes its maximal value. However, since the model $f$ is only an approximation of the true optimization landscape $F$, there is also incentive to explore other regions of the parameter space where few evaluations have been recorded. These two conflicting aspects, known as exploration-exploitation, are incorporated in Bayesian optimization by means of an acquisition function which values both the search for a maxima and also encourage exploration. The next set of parameters to evaluate is then chosen such that it maximizes this acquisition function. For example the upper confidence bound acquisition function, which was used in this work, is defined as
\begin{equation}\label{eq:bo:acq:ucb}
\alpha_{UCB}(\mathbf{x}) = \mu_f(\mathbf{x}) + k \sigma_f(\mathbf{x}) .
\end{equation}
$\mu_f(\mathbf{x})$ and $\sigma_f(\mathbf{x})$ are respectively the mean and the standard deviation of the predictive distribution given in Eq.~(\ref{update}). The parameter $k$ determines the exploration-exploitation balance with higher values of $k$ corresponding to more exploration. More details about the specificity of building and updating the probabilistic models within Bayesian optimization can be found in Refs.\cite{brochu2010tutorial, snoek2012practical, frazier2018tutorial, shahriari2015taking}.

\section*{References}
\bibliographystyle{unsrt} 
\bibliographystyle{iopart-num}

\end{document}